\documentclass[11pt,a4paper]{article}
\usepackage{jheppub}
\pdfoutput=1
\usepackage[utf8]{inputenc}
\usepackage{float}
\usepackage{lmodern}

\usepackage{amsmath,amssymb,mhchem}
\usepackage{graphicx}
\usepackage{url}
\usepackage{color}
\usepackage{enumitem}
\usepackage{subfigure}
\usepackage[dvipsnames]{xcolor}
\hypersetup{allcolors=[rgb]{0.0 0.0 0.6},linkcolor=[rgb]{0.75 0.05 0.05}}
\usepackage{bm}
\usepackage{epsfig}
\usepackage{amsmath}
\usepackage{amssymb}
\usepackage{slashed}
\usepackage{color}
\usepackage{accents}
\usepackage{url}
\usepackage{subcaption}
\usepackage[dvipsnames]{xcolor}
\usepackage{cancel}
\usepackage[normalem]{ulem}
\usepackage[colorinlistoftodos]{todonotes}
\usepackage{multirow}
\usepackage{booktabs} 
\usepackage[export]{adjustbox}

\newcommand{\exclude}[1]{}

\newcommand{\neut}{{\tilde{\chi}^0_1}}

\usepackage[normalem]{ulem}

\title{Probing light neutralinos from pair-produced sleptons with displaced vertices at the high-luminosity LHC}

\author[a,b]{Giovanna Cottin,}
\emailAdd{gfcottin@uc.cl}
\affiliation[a]{Instituto de F\'isica, Pontificia Universidad Cat\'olica de Chile, Avenida Vicu\~{n}a Mackenna 4860, Santiago, Chile}
\affiliation[b]{Millennium Institute for Subatomic Physics at the High Energy Frontier (SAPHIR), Fern\'andez Concha 700, Santiago, Chile}

\author[c,b]{Juan Carlos Helo,}
\emailAdd{jchelo@userena.cl}
\affiliation[c]{Departamento de F\'{i}sica,
Facultad de Ciencias, Universidad de La Serena, Avenida Cisternas 1200, La Serena, Chile}

\author[d,b]{Fabi\'an Hern\'andez-Pinto,}
\emailAdd{fabianhernandezpinto@gmail.com}
\affiliation[d]{Departamento de F\'isica, Universidad T\'ecnica Federico Santa Mar\'ia, Casilla 110-V, Valpara\'iso, Chile}

\author[e]{Nicol\'as A. Neill,}
\emailAdd{naneill@outlook.com}
\affiliation[e]{Centro Multidisciplinario de F\'isica, Vicerrector\'ia de Investigaci\'on, Universidad Mayor, 8580745 Santiago, Chile}

\author[f]{Zeren Simon Wang} 
\emailAdd{wzs@hfut.edu.cn}
\affiliation[f]{School of Physics, Hefei University of Technology, Hefei 230601, China}

\date{\today}

\abstract{We study light neutralinos ($\tilde \chi_1^0$) with masses ranging from 10 GeV to several hundred GeV within the framework of R-parity-violating (RPV) supersymmetry. These light neutralinos can be long-lived, decaying with a macroscopic displacement (order cm) inside the LHC main detectors. Complementing previous works on the subject, here we focus on their production through the electroweak pair production of left-chiral sleptons ($\tilde e_{L}$), with the signal process $pp\to \tilde e^+_{L} \tilde e^-_{L} \to e^+ \tilde \chi_1^0 e^- \tilde \chi_1^0$. In contrast to the previous study with a singly produced slepton, where the RPV coupling $\lambda'_{111}$ induces \textit{both} the production and decay of the light neutralino, in our scenario the production proceeds through Drell-Yan-like processes that are essentially independent of RPV couplings. Correspondingly, we implement a displaced-vertex search strategy for which our numerical analysis shows that the high-luminosity LHC can probe $\lambda'_{111}$ values up to three orders of magnitude smaller, and neutralino masses up to about four times larger than those accessible in the previously studied single-slepton production scenario.
}

\begin{document}

\maketitle
{
  \hypersetup{linkcolor=black}
}

\section{Introduction}

Interest in searching for R-parity-violating (RPV) interactions~\cite{Barbier:2004ez,Dreiner:1997uz,Mohapatra:2015fua} has grown significantly in recent years.
This is mainly because a priori the R-parity doesn't need to be conserved, and that RPV Supersymmetry (SUSY) in general offers a richer phenomenology~\cite{Dreiner:1991pe,deCampos:2007bn,Dercks:2017lfq,Dreiner:2023bvs,Dreiner:2025kfd} than the usual R-parity-conserving (RPC) SUSY.
In particular, the lightest supersymmetric particle (LSP) is no longer stable and can decay into Standard-Model (SM) particles.
RPV SUSY is well motivated for various reasons, including solving the issue of non-zero masses of active neutrinos~\cite{Hall:1983id,Grossman:1998py,Hirsch:2000ef,Dreiner:2006xw,Dreiner:2011ft}, explaining $B$-anomalies~\cite{Trifinopoulos:2019lyo,Hu:2020yvs,BhupalDev:2021ipu}, and addressing the muon $g-2$ anomaly~\cite{Hu:2019ahp,Zheng:2021wnu,BhupalDev:2021ipu}, in addition to the usual motivations of RPC SUSY, except that RPV SUSY does not offer a DM candidate in general.
At the LHC, collider signatures in the context of RPV SUSY include both prompt decays and those related to long-lived particles (LLPs).\footnote{See Refs.~\cite{Alimena:2019zri,Lee:2018pag,Curtin:2018mvb,Beacham:2019nyx} for recent overviews of LLP searches.}
For summaries of the current searches and limits for prompt signatures within RPV SUSY, we refer to Refs.~\cite{Dercks:2017lfq,Dreiner:2023bvs,Dreiner:2025kfd}. On the other hand, searches for long-lived neutralinos in RPV SUSY have both been reported by ATLAS and CMS~\cite{ATLAS:2015oan,CMS:2016vuw,ATLAS:2019fwx,ATLAS:2023oti} and studied in numerous phenomenological works (see, for instance, Refs.~\cite{Helo:2018qej,Dercks:2018eua,Dey:2020juy,deVries:2015mfw,Candia:2021bsl,Cottin:2022gmk}).

In this work, we focus on the lightest neutralino ($\neut$) with masses ranging from $10\,\text{GeV}$ to several hundred GeV.
We restrict ourselves to a bino-like neutralino~\cite{Gogoladze:2002xp,Dreiner:2009ic} which is supposed to be the LSP.
A bino as light as 10 GeV or even massless~\cite{Domingo:2022emr} is allowed by all laboratory~\cite{Dreiner:1991pe, Choudhury:1999tn, Dreiner:2009er,Dreiner:2009ic,Gogoladze:2002xp,Dreiner:2022swd} as well as astrophysical and cosmological constraints~\cite{Grifols:1988fw,Ellis:1988aa,Lau:1993vf,Dreiner:2003wh,Dreiner:2013tja,Profumo:2008yg,Dreiner:2011fp}, if the following conditions are satisfied: (1) the GUT-dictated relation between the gauginos $M_1\approx 0.5 M_2$ is not required~\cite{Choudhury:1995pj,Choudhury:1999tn}, (2) the light bino does not comprise the dark matter~\cite{Belanger:2002nr,Hooper:2002nq,Bottino:2002ry,Belanger:2003wb,AlbornozVasquez:2010nkq,Calibbi:2013poa}, and (3) the light bino is unstable so that the Universe is not overclosed~\cite{Bechtle:2015nua}.
In our case, the light bino decays via RPV couplings.
For a more detailed discussion, see e.g.~Ref.~\cite{deVries:2015mfw}.
Concretely, we will work with the Minimal Supersymmetric Standard Model (MSSM) extended with RPV interactions (RPV-MSSM).

In a previous work~\cite{Cottin:2022gmk}, we proposed a displaced-vertex (DV)-based search strategy at the high-luminosity LHC (HL-LHC), targeting a long-lived light neutralino with masses between 10 GeV and 230 GeV in the RPV-MSSM. The light neutralino is produced in the process $pp\to \tilde{e}_L\to e \neut$ via the RPV coupling $\lambda'_{111}$, and decays via the same coupling into an electron and two quark-jets.
In this case, the slepton is singly produced, and the RPV coupling mediates \textit{both} the production and decay of the light neutralino.
With the proposed search strategy, we have found that values of $\lambda'_{111}$ between $\sim \mathcal{O}(10^{-7})-\mathcal{O}(10^{-2})$ can be probed, reaching up to almost three orders of magnitude below the expected sensitivities in the single-production case~\cite{Cottin:2022gmk}.

In this article, we apply a similar search strategy for the HL-LHC, focusing instead on pair production of the sleptons which both decay promptly into an electron and a light neutralino.\footnote{Sensitivities of present and future LHC far detectors to this theoretical scenario have been studied in Refs.~\cite{Domingo:2023dew,Choudhury:2023yfg}. Further, similar search proposals for a long-lived right-handed neutrino in the context of a left-right symmetric model was performed in Refs.~\cite{Cottin:2019drg,Cottin:2018kmq}.}
Here, the production of a pair of the light neutralinos is induced by electroweak couplings in the MSSM, while the (displaced) decay of the light neutralinos proceeds via a non-vanishing RPV coupling $\lambda'_{111}$ mediated by off-shell sleptons.
See figure~\ref{fig:feyndiagrampairproduction} for the Feynman diagram illustrating the signal process.
Thus, the production and decay are decoupled, potentially allowing for stronger sensitivity reach to the $\lambda'_{111}$ coupling, provided that the pair-production rate is sufficiently large.

Although we focus throughout this work on the case with only $\lambda'_{111} \neq 0$, it is worth noting that the DV strategy we explore is, in principle, also sensitive to other couplings of the form $\lambda'_{1ij}$, such as $\lambda'_{112}$, $\lambda'_{121}$, and $\lambda'_{122}$.
This is because the neutralino decays into an electron and two quarks via an off-shell slepton, and the resulting jets can originate from various combinations of light quarks.
In the case of $\lambda'_{111}$, the final state involves $u$ and $d$ quarks; however, for $\lambda'_{112}$, $\lambda'_{121}$, and $\lambda'_{122}$, the jets involve $u$-$s$, $c$-$d$, and $c$-$s$ quark pairs, respectively.
Since we do not distinguish the jet flavor in our analysis and focus only on the presence of displaced tracks and electrons, our search strategy is equally applicable to all these $\lambda'_{1ij}$ couplings.

This paper is structured as follows.
In Sec.~\ref{sec:model} we describe the theoretical model of the RPV-MSSM as well as the benchmark scenario of the RPV coupling $\lambda'_{111}$.
We then elaborate on the proposed DV-based search strategy and explain the simulation procedure in Sec.~\ref{sec:simulation}, where we also present a cutflow and efficiency results.
In Sec.~\ref{sec:analysis}, we present the numerical results of the sensitivity reach of the HL-LHC to the RPV coupling $\lambda'_{111}$ as functions of the masses of the light neutralino and the slepton.
We conclude the work in Sec.~\ref{sec:conclusions}.
Additionally, in Appendix~\ref{app:checkmate}, we explain the procedure of reinterpreting an ATLAS search for a prompt dilepton and transverse missing  momentum or missing energy (MET), with the tool \texttt{CheckMATE2}~\cite{Dercks:2016npn}, in terms of our theoretical scenario.

\section{Model and benchmark scenario}\label{sec:model}

The RPV-MSSM corresponds to the RPC MSSM supplemented with the following superpotential terms that violate $R$-parity~\cite{Weinberg:1981wj,Hall:1983id}
\begin{equation}
    W_{\text{RPV}}
    = \sum_i \mu_i L_i H_u 
    + \sum_{i,j,k} \left(
      \frac{1}{2}\lambda_{ijk} L_i L_j E^c_k
      + \lambda'_{ijk} L_i Q_j D^c_k
      + \frac{1}{2}\lambda''_{ijk} U^c_i D^c_j D^c_k
    \right),\label{eq:WRPV}
\end{equation}
where $H_u,L_i,E_i^c,Q_i,U^c_i,D_i^c$ are chiral superfields and $i,j,k \in (1, 2, 3)$ generation indices.
This superpotential introduces 48 additional terms to the RPC MSSM.
The first three sets of operators violate lepton number ($L$) while the $\lambda^{\prime\prime}_{ijk}$ terms violate baryon number ($B$). Allowing both $L$-violating and $B$-violating terms to be non-vanishing would lead to too fast proton-decay rates into SM particles (unless the couplings are all extremely small), in conflict with current upper bounds.
Thus, in this work, we assume a $\mathbb{Z}_3$-symmetry called baryon triality ($B_3$)~\cite{Ibanez:1991pr,Dreiner:2012ae} so that the $\lambda''$ couplings are vanishing and proton decays would not be induced.
The $L$-violating terms are constrained by the non-observation of neutrinoless double beta ($0\nu\beta\beta$) decay~\cite{Hirsch:1995ek,Bolton:2021hje} or other low-energy processes~\cite{Allanach:1999ic}.

In this work, we choose to focus on the operator $L_i Q_j D^c_k$ and the corresponding Yukawa couplings generated by this operator are 
\begin{align}
    L_{\text{RPV}} = \lambda'_{ijk}
    \left(
      \tilde \nu_{i L} \bar d_{k R} d_{j L}
      + \tilde d_{j L} \bar d_{k R} \nu_{i L}
      + \tilde d^*_{k R} \bar \nu^c_{iR} d_{j L}
      - \tilde e_{i L} \bar d_{kR} u_{j L}
    \right.\ \ \ \nonumber\\
    \left.
      - \tilde u_{j L} \bar d_{kR} e_{i L}
      - \tilde d^*_{k R} \bar e^c_{iR} u_{j L}
    \right) + \text{h.c.}\label{eq:LRPV}
\end{align}

For simplicity, we assume that all superpartners other than $\tilde \chi_1^0$ and the lightest slepton ($\tilde e_{L}$) are heavy (10 TeV) and therefore decoupled for the phenomenology.
Additionally, we assume that in the RPV Lagrangian, Eq.~\eqref{eq:LRPV}, only the coupling $\lambda'_{111}$ is nonzero.
Consequently, the phenomenology is controlled by the following parameters:
\begin{equation}
\lambda'_{111}, m_{\tilde e_L}, m_{\tilde \chi^0_1}.
\end{equation}

\begin{figure}
    \centering
    \includegraphics[width=0.5\textwidth]{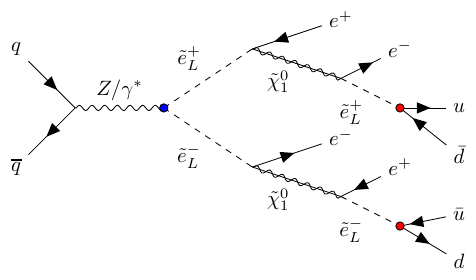}
    \caption{Slepton pair production in Drell-Yan-like processes (blue vertex) and subsequent decay of each slepton into a prompt lepton and a light neutralino.
The neutralinos are long-lived since they can only decay through the suppressed RPV coupling $\lambda^\prime_{111}$ (red vertices).}
    \label{fig:feyndiagrampairproduction}
\end{figure}

In this scenario, the lightest sleptons ($\tilde e_{L}^{\pm}$) can be pair-produced at the LHC in a Drell-Yan-like process, subsequently decaying to a prompt charged lepton and a neutralino (see figure~\ref{fig:feyndiagrampairproduction}).
Since the neutralino $\tilde{\chi}^0_1$ is the LSP, it can only decay through the RPV coupling $\lambda^\prime_{111}$, cf.~Eq.~\eqref{eq:LRPV}.
The total decay width entails the following relation with respect to $\lambda'_{111}, m_{\tilde e_L}, \text{ and } m_{\tilde \chi^0_1}$,
\begin{equation}
    \Gamma_{\tilde \chi^0_1} \propto m_{\tilde \chi^0_1}^5\left(\frac{\lambda_{111}^{'}}{m_{\tilde e_L}^2}\right)^2.
    \label{eq:Gamma}
\end{equation}
As the RPV operator $LQD^c$ violates $L$, the coupling $\lambda^{\prime}_{111}$ is tightly constrained by the non-observation of neutrinoless double beta ($0\nu\beta\beta$) decays at experiments~\cite{Hirsch:1995ek,Bolton:2021hje}, excluding $\lambda^{'}_{111}$ values larger than $\mathcal{O}(10^{-2})$ across GeV-scale neutralino masses, rendering GeV-scale $\neut$ naturally long-lived.

In the previous study of this scenario, we have focused on the process $pp\to \tilde e_L \to e \tilde \chi_1^0$~\cite{Cottin:2022gmk}, where the neutralino production cross section depends on $\lambda'_{111}$ and is therefore suppressed by the smallness of the RPV couplings.
In contrast, in the present work, the neutralinos originate from the electroweak pair production of the sleptons, and consequently their production cross section is independent of the RPV couplings and thus only the neutralino decay width is suppressed by the RPV coupling $\lambda'_{111}$.

\begin{figure}[t]
    \centering
    \includegraphics[scale=0.9]{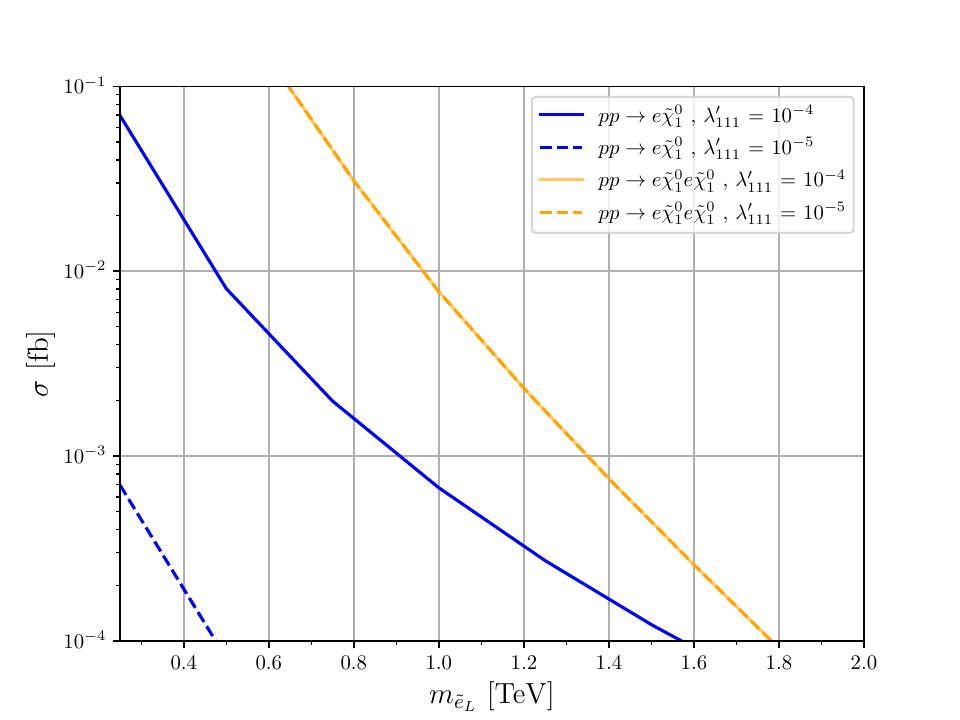}
    \caption{Production cross section for the single (blue) and pair (orange) production of neutralinos as a function of slepton mass. The neutralino mass is fixed at $m_{\tilde \chi_1^0}=100$ GeV. Neutralino single production was studied in Ref.~\cite{Cottin:2022gmk}.}
    \label{fig:xsec}
\end{figure}

In our study, we will present our results considering (i) fixed values of the slepton mass, $m_{\tilde e_L} = (0.5, 0.75, 1,1.25)~\text{TeV}$ while we vary freely the other parameters ($\lambda'_{111}$ and $m_{\tilde \chi_1^0}$) following the same practice as done in Ref.~\cite{Cottin:2022gmk}; and (ii) fixed values of $\lambda^{\prime}_{111}$ while varying freely the masses ($m_{\tilde e_L}$ and $m_{\tilde \chi_1^0}$).
Figure \ref{fig:xsec} shows the cross sections of the neutralino pair production, $pp\to \tilde e_{L}^+ \tilde e_{L}^- \to e^+ \tilde \chi_1^0e^-\tilde \chi_1^0$, and neutralino single production, $pp\to \tilde e_L \to e\tilde \chi_1^0$; the latter is extracted from Ref.~\cite{Cottin:2022gmk}.
The fact that the neutralino pair production cross section is independent of the RPV couplings\footnote{This is only valid for values of $\lambda'_{111}$ that are sufficiently small, as the narrow-width approximation is inaccurate for $\lambda'_{111}\gtrsim 10^{-1}$ (see also the erratum of Ref.~\cite{Cottin:2022gmk}).} will allow us to probe smaller values of $\lambda'_{111}$ compared to the scenario with singly produced neutralinos.

\section{Simulation and event selection}\label{sec:simulation}

We follow Ref.~\cite{Cottin:2022gmk} to implement a DV-based search strategy, with at least one DV signature emerging from the neutralino decay products (charged-particle tracks).
We use the RPV-MSSM UFO model implemented in Ref.~{\cite{RPVUFO}}, with flavor-diagonal couplings, and consider a bino-like neutralino.
We simulate the signal process $pp\to \tilde e^{+}_{L} \tilde e^{-}_{L}  \to e^{+}\tilde \chi_1^0 e^{-}\tilde \chi_1^0$~\footnote{Note that, as mentioned in the previous section, in our scenario all the supersymmetric particles other than $\tilde \chi_1^0$ and $\tilde e_L$, including $\tilde e_R$, are assume to be heavy and therefore decoupled from the phenomenology.} in \texttt{MadGraph5}~\cite{Alwall:2011uj} at $\sqrt{s}=13$ TeV,\footnote{Default kinematic cuts for the outgoing electron or positron ($p_{T} \geq 10$ GeV and $|\eta|<2.5$) are used in \texttt{Madgraph5} at the generator level.} and generate parton-level LHE events with displaced information (with the \texttt{time\_of\_flight} flag switched on).
We consider the decay of each selectron ($\tilde e_L$) into a prompt electron and the lightest neutralino ($\tilde \chi^0_1$).
The decay widths of $\tilde e_L$ and $\tilde \chi_1^0$ are automatically computed by \texttt{MadGraph5}.
We use \texttt{Pythia8}~\cite{Sjostrand:2014zea} to perform showering and hadronization on the LHE files from \texttt{MadGraph5}, with a custom-made code implemented within \texttt{Pythia8} that includes the cuts shown in Table~\ref{tab:DVcuts} for the signal analysis, inspired by an ATLAS search~\cite{Aaboud:2017iio}.
\begin{table}[t]
\centering
\begin{tabular}{ll}
\toprule
\midrule
Trigger selection      & $\cdot$  A reconstructed electron with $p_{T}>25$ GeV and with $|\eta| < 2.47$           \\
DV selection           & $\cdot$ Displaced tracks with $|d_{0}|> 2$ mm and $p_{T} > 1$ GeV \\
                       & $\cdot$ Transverse decay position of the DV, $r_{\text{DV}}$ between 4 and 300 mm  \\
                       & $\cdot$ Longitudinal decay position of the DV $|z_{\text{DV}}| < 300$ mm. \\
                       & $\cdot$ Number of charged tracks coming from the DV $n_{\text{trk}} \geq 5$ \\
                       & $\cdot$ Invariant mass of the DV $m_{\text{DV}} \geq 10$ GeV\\
\bottomrule
\end{tabular}
\caption{Selections for the displaced vertex (DV) analysis. Here, $d_0$ labels transverse impact parameter. }
\label{tab:DVcuts}
\end{table}
Electrons are reconstructed from Monte-Carlo truth information with momenta smearing as in Ref.~\cite{Cottin:2022gmk} and within $|\eta|<2.47$.
Our event selection starts with a trigger cut of 25 GeV on the transverse momentum of the reconstructed electrons.
We then proceed to reconstruct displaced vertices from displaced tracks.
The last two DV selections in Table~\ref{tab:DVcuts} highlight the signal region~\cite{Aaboud:2017iio} where there are no Standard Model backgrounds\footnote{After these last two cuts, backgrounds for displaced-vertex searches are low ($\sim 0.02$ background events reported in the ATLAS search in Ref.~\cite{Aaboud:2017iio} for an integrated luminosity of 32.8 fb$^{-1}$) and purely instrumental in origin. Given the limitations of estimating them outside the experimental collaborations, we assume zero background in our analysis. In addition, for the purpose of our estimations in section~\ref{sec:analysis}, we assume the DV efficiencies will remain the same at higher luminosities. } .

We make use of the parameterized vertex-level efficiencies~\cite{Aaboud:2017iio} in the $m_{\text{DV}}$ vs.~$n_{\text{trk}}$ plane to quantify the ATLAS detector response to the displaced vertices. 

Three benchmark scenarios were selected to cover different regimes of the proper decay length of $\neut$, $c\tau$: B1 corresponds to a promptly-decaying neutralino ($c\tau$ is less than a millimeter), B2 to a neutralino that predominantly decays inside the inner detector (between 4 and 300 mm), while B3 to a neutralino with $c\tau$ of the order of meters.
The cutflows for these benchmarks are summarized in Table~\ref{tab:DVcutflows}.
Among them, the benchmark B2 is the most efficient, as its $c\tau$ falls within the bulk of the acceptance for a DV.
\begin{table}[t]
    \centering
    $c\tau \approx$ 0.59 mm \quad $\lambda^{\prime}_{111}$ = 10$^{-3}$ \\
    \begin{tabular}{c|c|c|c}
        \hline
    \toprule
     B1 &  \textbf{Number of events} & Relative \% & Total \% \\
    \midrule
    All events  & 10000 & 100 & 100 \\
     $e^\pm$ Trigger & 9967 & 99.67 & 99.67 \\
    DV  Fiducial & 1307 & 13.11 & 13.07 \\
    DV $n_{\text{trk}}$ & 331 & 25.32 & 3.31 \\
    DV Mass & 217 & 65.55 & 2.17 \\
    DV Eff & 157 & 72.35 & 1.57 \\
    \bottomrule
\end{tabular}    

            $c\tau \approx$ 59 mm \quad $\lambda^{\prime}_{111}$ = 10$^{-4}$ \\
    \begin{tabular}{c|c|c|c}
    		\hline
        \toprule
     B2 &  \textbf{Number of events} & Relative \% & Total \% \\
        \midrule
        All events  & 10000 & 100 & 100 \\
         $e^\pm$ Trigger & 9979 & 99.79 & 99.79 \\
        DV  Fiducial & 7408 & 74.23 & 74.08 \\
        DV $n_{\text{trk}}$ & 6896 & 93.08 & 68.96 \\
        DV Mass & 6780 & 98.31 & 67.8 \\
        DV Eff & 3120 & 46.01 & 31.2 \\
        \bottomrule
    \end{tabular}\\
              $c\tau \approx$ 5.9 m \quad $\lambda^{\prime}_{111}$ = 10$^{-5}$ \\
    \begin{tabular}{c|c|c|c}
    		\hline
        \toprule
         B3 &  \textbf{Number of events} & Relative \% & Total \% \\
        \midrule
        All events  & 10000 & 100 & 100 \\
        $e^\pm$ Trigger& 9802 & 98.02 & 98.02 \\
        DV Fiducial & 211 & 2.15 & 2.11 \\
        DV $n_{\text{trk}}$ & 203 & 96.20  & 2.03 \\
        DV Mass & 200 & 98.52 & 2 \\
        DV Eff &  62  & 31 & 0.62 \\
        \bottomrule
    \end{tabular}\\     
   \caption{Cutflows for the defined benchmarks B1, B2, and B3. All the benchmarks have a fixed value of $m_{\tilde{\chi}^0_1}$ and $m_{\tilde{e}_L}$ equal to 100 GeV and 0.5 TeV, respectively.}
    \label{tab:DVcutflows}
\end{table}

\begin{figure}[t]
	\centering
	\includegraphics[scale=1]{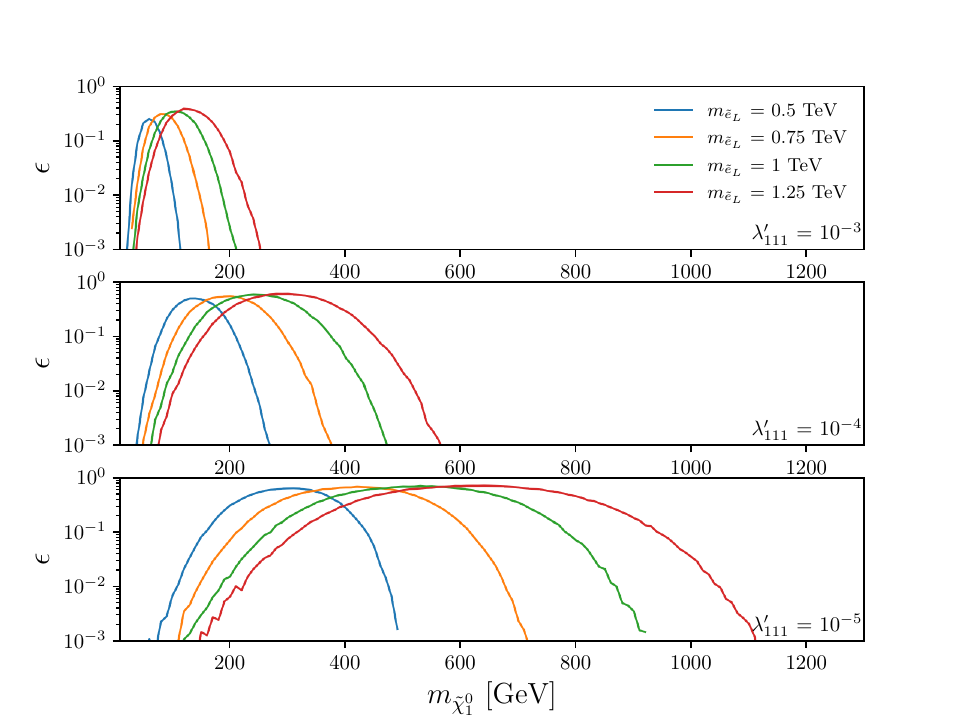}
	\caption{DV selection efficiencies as functions of the neutralino mass, for fixed choices of $\lambda^{'}_{111}$ and $m_{\tilde{e}_L}$.}
	\label{fig:DVeff}
\end{figure}

In figure \ref{fig:DVeff}, we show the overall selection efficiencies of our DV strategy as functions of $m_{\tilde \chi_1^0}$ for several values of $m_{\tilde{e}_L}$ and $\lambda^{\prime}_{111}$. The range of neutralino masses with a sizable efficiency is wider for lower values of $\lambda^{\prime}_{111}$. This can be understood by the scaling behavior of $\Gamma_{\neut}$ with respect to $m_{\neut}$ and $\lambda'_{111}$, cf.~Eq.~\eqref{eq:Gamma}. The peak in efficiency is reached for combinations of the masses and couplings yielding similar values of $c\tau$. The highest efficiency is obtained for values of the boosted decay length of the neutralino,
\begin{equation}
\beta\gamma c\tau \propto \frac{|\vec p_{\chi_1^0}|}{\lambda^{\prime 2}_{111}}\frac{m^4_{\tilde{e}_L}}{m^6_{\tilde{\chi}^0_1}},
\end{equation}
of the order $\mathcal{O}(\text{cm})$, which corresponds to the ATLAS inner detector size.

\begin{figure}[t]
	\centering
	\includegraphics[scale=1]{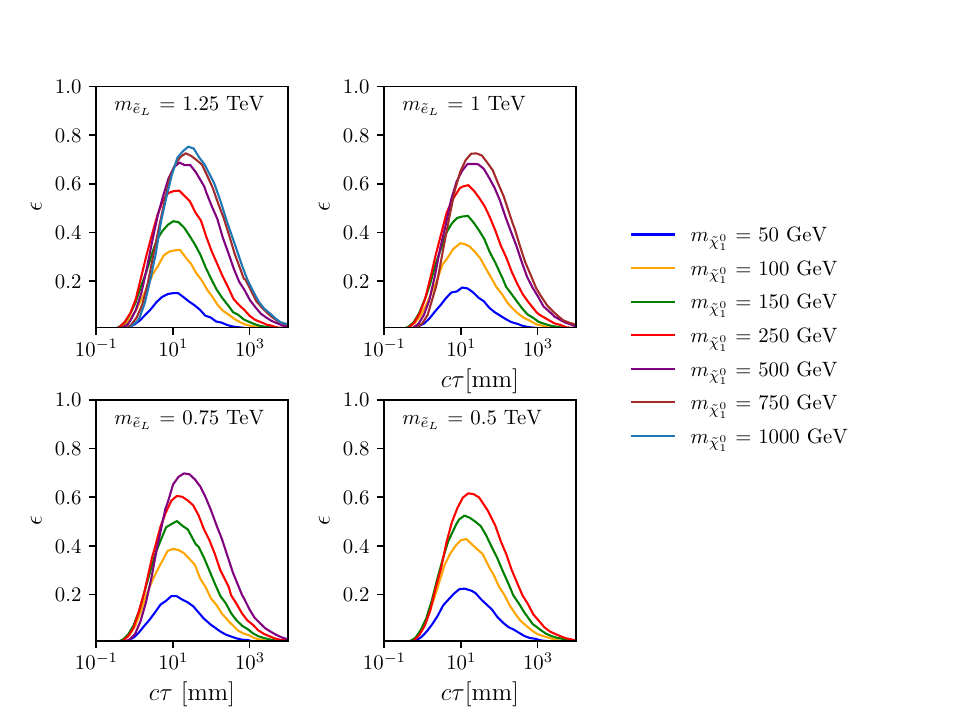}
	\caption{DV selection efficiency as a function of the $c\tau$ of the neutralino, for fixed choices of $m_{\tilde{e}_L}$ and $m_{\tilde{\chi}^{0}_{1}}$.}
	\label{fig:DVvsctau}
\end{figure}

In Figure~\ref{fig:DVvsctau}, we show the DV efficiencies as functions of the neutralino proper decay length, for different slepton and neutralino masses. It is clear that the efficiencies are enhanced for larger neutralino masses, independently of the selectron mass. This is because higher neutralino masses result in more available tracks that contribute to $m_{\text{DV}}$.

\section{LHC sensitivity with displaced-vertex searches}\label{sec:analysis}

\subsection{Sensitivity to $\lambda^{\prime}_{111}$}

We proceed to compute the sensitivity reach of the ATLAS inner tracker detector to the trilinear RPV coupling $\lambda'_{111}$ with our DV search strategy, for long-lived light neutralinos.

\begin{figure}[t]
\centering
\includegraphics[scale=0.5]{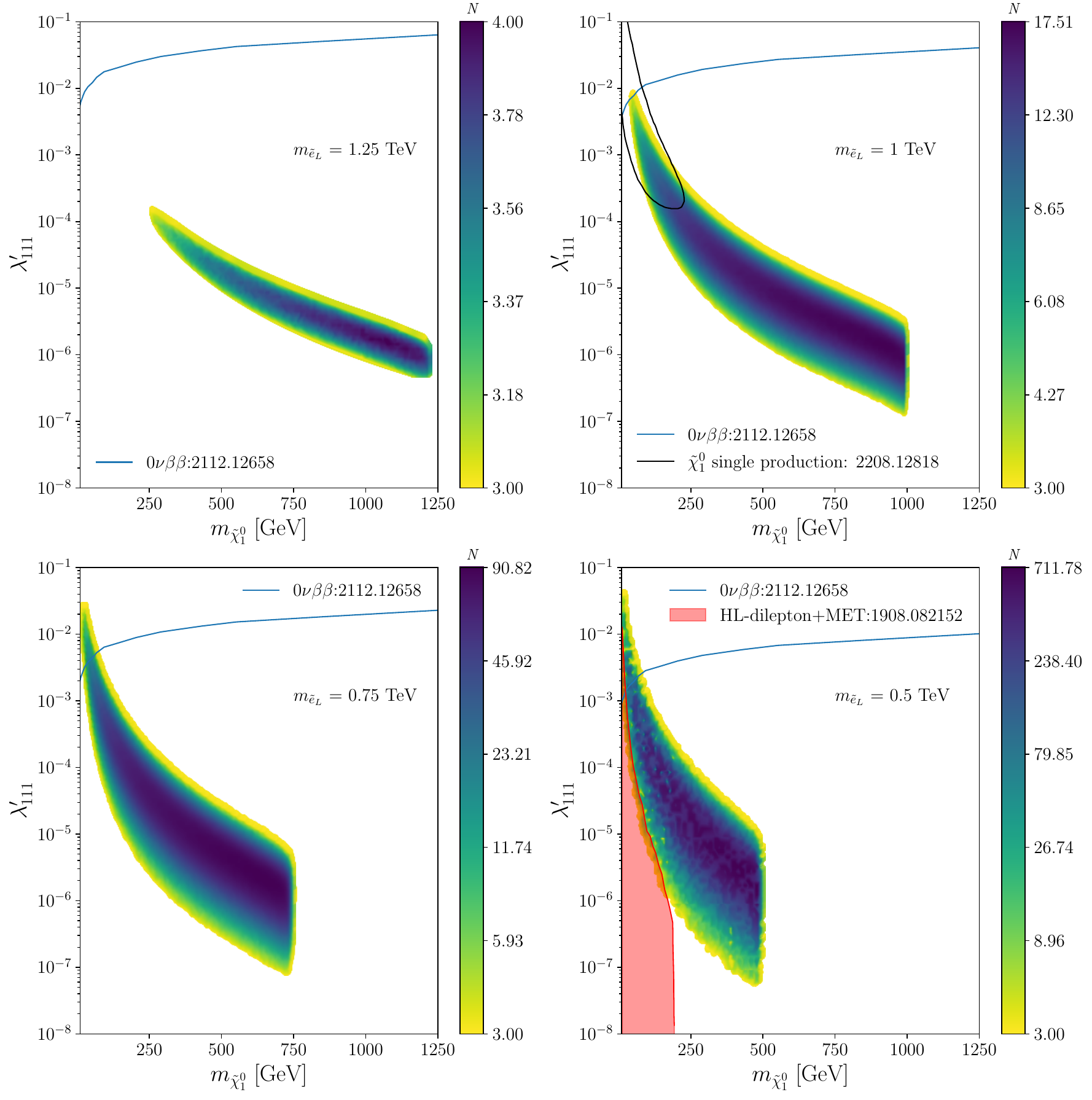}
\caption{Sensitivity reach in the $\lambda'_{111}$ vs.~$m_{\tilde \chi_1^0}$ plane with the proposed DV search strategy for an integrated luminosity of $\mathcal{L}=$ 3000 fb$^{-1}$. The number of expected signal events depicted in the color bar starts from 3. The red shaded area corresponds to the parameter region excluded by the ``2-leptons + MET'' search recast implemented in \texttt{CheckMATE2} at $\mathcal{L}=$ 3000 fb$^{-1}$. The blue lines correspond to the limits from searches for $0\nu\beta\beta$ decays at GERDA~\cite{Bolton:2021hje}. In the upper right plot ($m_{\tilde{e}_L}=1$ TeV), we overlap our results with those from the study on the long-lived lightest neutralino with the single production of the slepton~\cite{Cottin:2022gmk}.}
\label{fig:limits_final}
\end{figure}

Figure~\ref{fig:limits_final} shows the sensitivity reach in the $\lambda^{\prime}_{111}$vs.~$m_{\tilde{\chi}^0_1}$ plane for 4 selectron masses (1.25 TeV, 1 TeV, 0.75 TeV, and 0.5 TeV).
The border of the colored region corresponds to the 95\% C.L.~exclusion bounds under the zero-background assumption\footnote{We note that new experimental techniques could keep potentially non-negligible backgrounds under control at higher luminosities, e.g.~improved architectures in secondary-vertex reconstruction~\cite{Shlomi:2020ufi} or novel displaced triggers~\cite{Alimena:2021mdu}. Nevertheless, as we have displayed exclusion contours of both $N=3$ and higher $N$-values in figure~\ref{fig:limits_final}, this presentation should help understand better the effect of smaller efficiencies or higher levels of background.}. We also overlay our results with constraints from a prompt search for a dilepton + MET, recast and reinterpreted with CheckMATE2~\cite{Dercks:2016npn}.
The details of this reinterpretation are presented in Appendix~\ref{app:checkmate}.
We also display the limits from searches for $0\nu\beta\beta$ decays at GERDA~\cite{Bolton:2021hje} for completeness.
Finally, for the case of $m_{\tilde{e}_L}=1$ TeV, we overlay for comparison the corresponding bounds from the single production of $\tilde{e}_L$ obtained in Ref.~\cite{Cottin:2022gmk}.

In figure~\ref{fig:limits_final} we first observe that, in general, the expected number of signal events drops down with increasing mass of the selectron.
This is explained by the fact that at higher slepton masses the production cross section decreases down to the order of $\mathcal{O}(10^{-3})$ fb, close to the limit value that allows us to have nonzero sensitivity.
The proposed DV strategy can reach $\lambda^{\prime}_{111}$ values of $\mathcal{O}(10^{-7})$ for a wide range of neutralino masses.
The strongest sensitivity is expected for the case of $m_{\tilde{e}_L}$ = 0.5 TeV, probing $\lambda^{\prime}_{111}$ between 10$^{-7}$ -- 10$^{-1}$ and $m_{\tilde{\chi}^0_1} $ from 10 GeV up to about $m_{\tilde{e}_L}$. These sensitivity results are shown to largely exceed the bounds from $0\nu\beta\beta$-decay considerations. 
The shape of the sensitivity region is bounded by boosted neutralinos that decay too promptly (upper part) or escape the outer boundaries of the ATLAS inner tracker (lower part).
In contrast to the single-production case~\cite{Cottin:2022gmk}, sensitivities to the neutralino masses very close (but not equal) to the selectron mass are achieved.
This is because there is no cross-section dependence on $\lambda^{\prime}_{111}$, in the small $\lambda^{\prime}_{111}$ limit; the sensitivities do not extend beyond this kinematic threshold because the off-shell contributions are negligible.
In addition, an interesting effect is worth noticing.
Since we do not place any cuts on the electron displacement at the trigger level, events can be triggered by not only prompt electrons but also \textit{displaced} ones originating from the neutralino decays.
This fact aids the sensitivity of our search, allowing to probe mass splittings between the selectron and the neutralino as low as 4 GeV (which is below the 25-GeV $p_{T}$ trigger cut on electrons).

The red area in the lower right plot corresponds to the parameter region excluded by our reinterpretation of the ``2-lepton + MET'' ATLAS search~\cite{ATLAS:2019lff} in the context of our RPV signal, in the framework of \texttt{CheckMATE2}.
The reinterpretation is performed for an integrated luminosity of  $\mathcal{L}=$ 3000 fb$^{-1}$. The points excluded by the search correspond to neutralinos with a $c\tau$ large enough to be considered fully stable, contributing to MET.
Given the large MET in this limit, the dilepton + MET search is very efficient, and falls off harshly at  $m_{\neut}\sim 190$ GeV for $m_{\tilde{e}_L}=0.5$ TeV.
This is due to the signal regions (SRs) chosen in the search, which place a cut on the $m_{T2}$ variable (see Appendix~\ref{app:checkmate} for more details).

\subsection{Exclusion limits in the plane spanned by the slepton- and neutralino-masses}

\begin{figure}[t]
	\centering
	\begin{minipage}{0.49\textwidth}
		\centering
		\includegraphics[width=\linewidth]{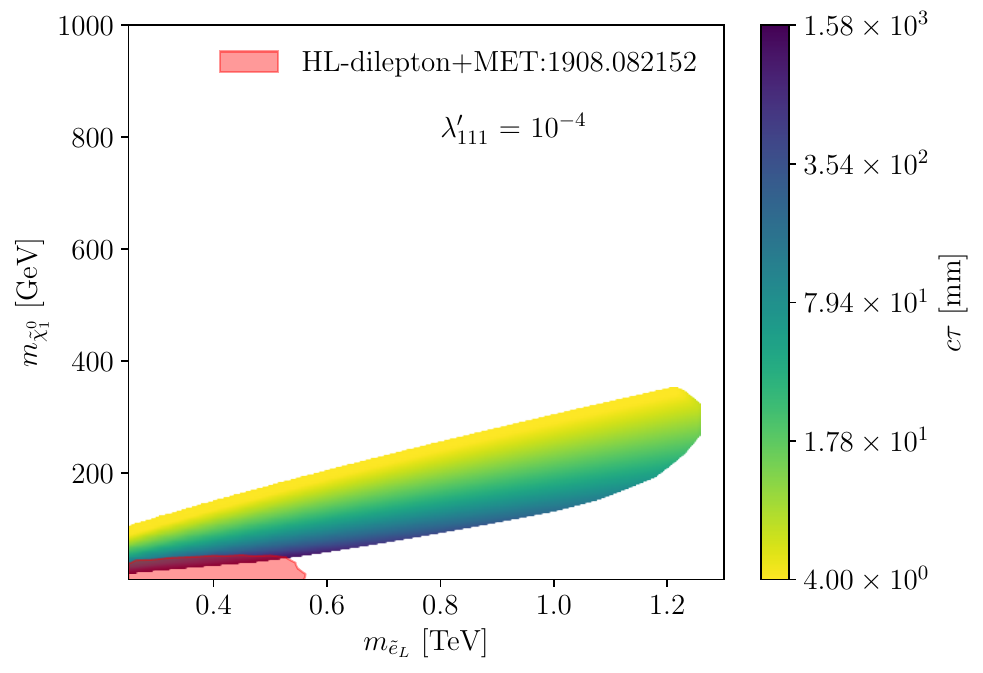}
	\end{minipage}
	\hfill
	\begin{minipage}{0.49\textwidth}
		\centering
		\includegraphics[width=\linewidth]{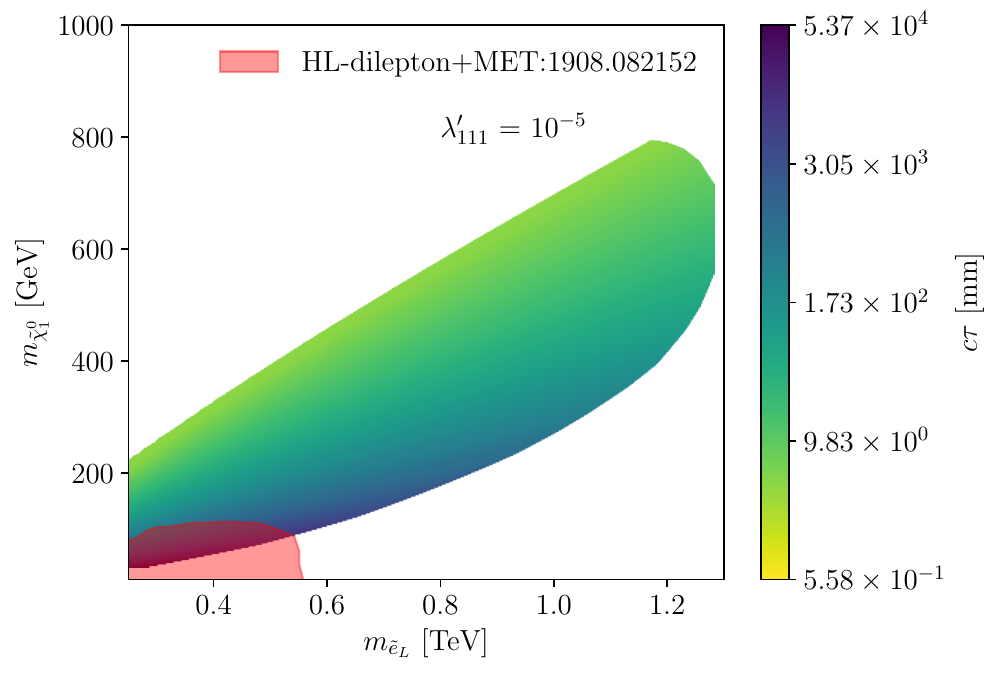}
	\end{minipage}
	
	\vskip 1em
	
	\begin{minipage}{0.5\textwidth}
		\centering
		\includegraphics[width=\linewidth]{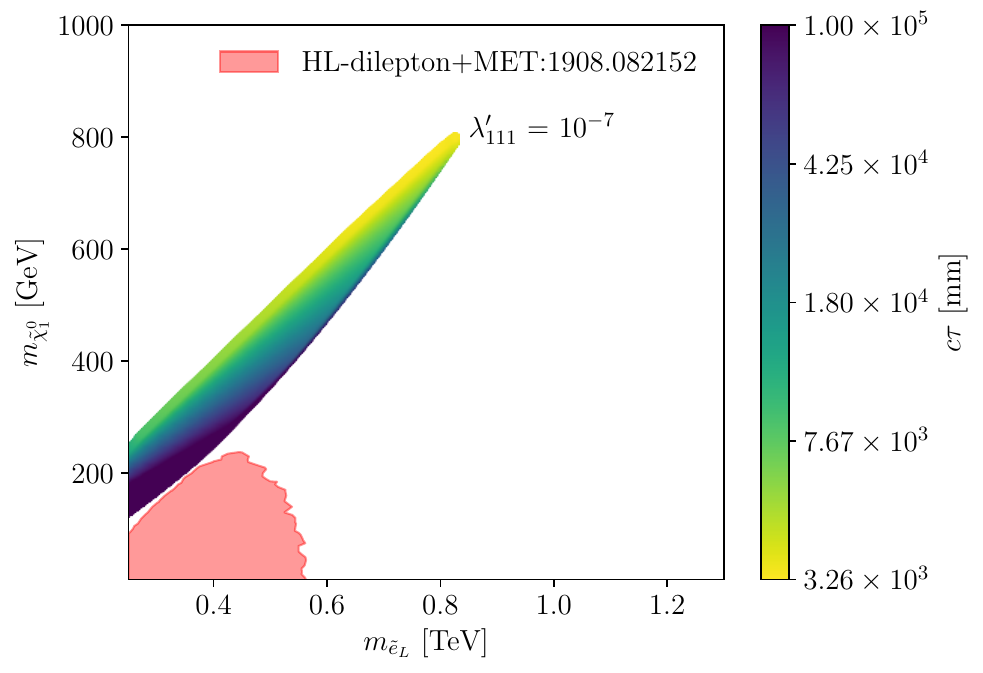}
	\end{minipage}
	
	\caption{DV sensitivity in the $m_{\neut}$ vs.~$m_{\tilde{e}_L}$ plane for an integrated luminosity of 3000 fb$^{-1}$. The color bar indicates the $c\tau$ values that respect $N\geq3$ for a given combination of selectron mass and neutralino mass, for a fixed $\lambda'_{111}$ coupling. On the verge of each sensitivity region, $N=3$ is predicted corresponding to exclusion limits at 95\% C.L.}      
	\label{fig:DVvsmass}
\end{figure}

In addition, we present sensitivity results in figure~\ref{fig:DVvsmass} in the $m_{\neut}$ vs.~$m_{\tilde{e}_L}$ plane, for fixed values of $\lambda'_{111}=10^{-4}, 10^{-5}$, and $10^{-7}$.
We observe that the two cases of relatively large values of $\lambda'_{111}$ can reach $m_{\tilde{e}_L}$ beyond 1.2 TeV while the case of $\lambda'_{111}=10^{-7}$ is sensitive to $m_{\tilde{e}_L}$ only up to about 0.85 TeV. Also, figure~\ref{fig:DVvsmass} shows that in the case of a large value of $\lambda'_{111}$ our proposed analysis is found to have a smaller upper reach to $m_{\neut}$.
For instance, for $\lambda'_{111}=10^{-4}$, the proposed search can constrain $m_{\tilde{\chi}^0_1}$ up to about 300 GeV, while for $\lambda'_{111}=10^{-7}$ it is sensitive to $m_{\tilde{\chi}^0_1}$ as large as $\sim 800$ GeV.

In figure~\ref{fig:DVvsmass}, the red area corresponds to the parameter region excluded by the dilepton + MET ATLAS search (with an integrated luminosity of 3000 fb$^{-1}$); this region corresponds to the limit of very long-lived neutralinos that could be considered as ``stable" and appearing as MET. However, if $m_{\neut}$ increases, $c\tau$ is lowered, resulting in less MET; as a result, the ATLAS dilepton + MET search becomes less efficient. If higher slepton masses are considered, the light-neutralino production cross section decreases, leading to fewer expected signal events.

\section{Conclusions}\label{sec:conclusions}

In SUSY models, sleptons can be pair produced and decay into a charged lepton and a light neutralino.
If we assume R-parity violation and that the lightest neutralino is both the LSP and bino-like, the light bino can decay via RPV couplings into SM particles.
Here, we consider a light bino in the RPV-MSSM with masses ranging from 10 GeV to around 1 TeV, produced in pairs via slepton decays and subsequently decaying via a single RPV coupling $\lambda'_{111}$.
With this coupling, we restrict ourselves to studying electrons and selectrons for the charged leptons and sleptons (sneutrinos and squarks are assumed to be sufficiently heavy to be decoupled from the phenomenology).
The light bino in the mass range of our interest can easily be long-lived if the RPV coupling is sufficiently small, and decay within the acceptance of the trackers at the LHC main detectors.

In this paper, we have proposed a DV-based search strategy at the HL-LHC, inspired by a previous work~\cite{Cottin:2022gmk}, which explored light long-lived neutralinos whose production and decay are both mediated by the same RPV coupling $\lambda'_{111}$.
In contrast, our study focuses on pair-produced light neutralinos, where the HL-LHC can probe smaller values of $\lambda'_{111}$. This improved sensitivity arises because the production cross section is independent of $\lambda'_{111}$, as long as the coupling is small enough not to significantly affect the slepton's total decay width.

We have performed Monte-Carlo simulations of our signal process, analyzed the acceptance of our proposed search analysis, and computed the sensitivity of the ATLAS detector at the HL-LHC to the coupling $\lambda'_{111}$ as functions of the masses of the selectron and the light neutralino.
The pair production of the selectrons is induced with electroweak couplings and the selectrons are assumed to be on-shell.
They decay promptly into an $e^-$ or $e^+$, plus a light bino.
The light bino is long-lived and displaced-decays into an $e^\pm$ and two quark-jets mediated by an off-shell selectron.

Besides the sensitivity reach of our proposed search, we have recast an ATLAS prompt search for a pair of leptons plus MET and reinterpreted it in terms of our light-bino scenario. Moreover, present bounds from searches for $0\nu\beta\beta$ decays are considered. We overlap the results of these searches when presenting the final sensitivities.

As in the single-production scenario studied in Ref.~\cite{Cottin:2022gmk}, the phenomenological scenario is controlled by three free parameters: $\lambda^{\prime}_{111}$, $m_{\tilde{e}_L}$ and $m_{\tilde{\chi}^0_1}$. 
We present numerical efficiencies as functions of the neutralino mass and proper decay length $c\tau$ in figure~\ref{fig:DVeff} and figure~\ref{fig:DVvsctau}.

Further, in figure~\ref{fig:limits_final} and figure~\ref{fig:DVvsmass} we show the final exclusion bounds at $95\%$ C.L.~in the planes $\lambda^{\prime}_{111}$ vs.~$m_{\tilde{\chi}^0_1}$ and $m_{\tilde{\chi}^0_1}$ vs.~$m_{\tilde{e}_L}$, with fixed values of $m_{\tilde{e}_L} =$ 1.25, 1, 0.75, and 0.5 TeV and fixed values of $\lambda^{\prime}_{111}$ = $10^{-4}$, $10^{-5}$, and $10^{-7}$, respectively. 

Figure~\ref{fig:limits_final} displays stronger sensitivity reach to $\lambda'_{111}$ for lighter selectrons, and that the HL-LHC can probe the bino mass up to $m_{\neut}\lesssim m_{\tilde{e}_L}$.
In addition, the sensitivity reach of our proposed strategy is complementary to that of the ongoing prompt dilepton + MET searches and is stronger than the $0\nu\beta\beta$-decay bounds by orders of magnitude.

In the case of $\lambda'_{111}=10^{-4}$ in figure~\ref{fig:DVvsmass}, we find that the HL-LHC can probe light bino masses in the range of approximately 20 GeV to 300 GeV, for selectron masses between 0.25 TeV and $\sim$1.25 TeV.
For $\lambda'_{111}=10^{-5}$, our proposed search strategy can reach neutralino masses up to about 800 GeV and slepton masses up to around 1.25 TeV at the HL-LHC.
In the case of $\lambda'_{111}=10^{-7}$, neutralino masses between about 150 GeV and 800 GeV and selectron masses from 0.25 TeV to 0.8 TeV, can be probed.
Finally, we find no sensitivity for $\lambda'_{111}\lesssim 10^{-8}$, which is consistent with the results presented in figure~\ref{fig:limits_final}.

Although our numerical results are based on the benchmark scenario where only $\lambda'_{111} \neq 0$, the DV search strategy we propose is, in principle, also sensitive to other couplings of the form $\lambda'_{1ij}$ with $i,j,=1,2$.
This is because in all these cases the neutralino decays into an electron and two jets;  in the additional cases, the jets originate from different combinations of light quarks that are experimentally indistinguishable with our strategy from those produced in $\lambda'_{111}$ decays. Therefore, our analysis can be interpreted as representative of a broader class of RPV couplings with similar experimental signatures.

To conclude, we find that the HL-LHC should be highly sensitive to long-lived light neutralinos pair-produced from slepton decays and decaying via RPV couplings into SM particles, as large parameter regions defined by $\lambda'_{111}, m_{\neut}$, and $m_{\tilde{e}_L}$ can be probed.
Moreover, a DV-based search strategy can be complementary to prompt searches and $0\nu\beta\beta$ searches, for constraining such LLP scenarios.

\acknowledgments

We thank Abi Soffer for useful discussions.
G.C. acknowledges support from ANID FONDECYT grant No. 1250135. G.C., F.H.P., and J.C.H. also acknowledge support from ANID FONDECYT grant No. 1201673, and J.C.H. from ANID FONDECYT grant No. 1241685. All three authors also acknowledge support from the ANID Millennium Science Initiative Program ICN2019\_044.
N.A.N. was supported by ANID (Chile) FONDECYT Iniciaci\'on
Grant No. 11230879.
Z.~S.~Wang was supported by the National Natural Science Foundation of China under Grant No.~12475106 and the Fundamental Research Funds for the Central Universities under Grant No.~JZ2025HGTG0252.

\appendix
\section{Recast of the ATLAS prompt dilepton + MET search with CheckMATE}
\label{app:checkmate}

\texttt{CheckMATE2}~\cite{Dercks:2016npn} includes a large library of experimental analyses from both ATLAS and CMS at various center-of-mass energies and with different integrated luminosities.
Also, \texttt{CheckMATE2} can load \texttt{Madgraph5\_aMC@NLO} and \texttt{Pythia8} directly, which allows for the generation of parton-level events within the same program workflow and their subsequent hadronization.
These events are then analyzed using \texttt{Delphes3}~\cite{deFavereau:2013fsa} for detector simulation and the results are passed to \texttt{ROOT} and prepared for event-by-event analysis.

For our RPV signal, we calculate a prompt limit based on the ATLAS analysis reported in Ref.~\cite{ATLAS:2019lff} which searches for pair production of neutralinos and charginos at the electroweak scale in R-parity-conserving SUSY, which is implemented in \texttt{CheckMATE2}.
This search matches our signature for a stable neutralino, where the neutralino is produced in pairs together with two leptons. The experimental signature corresponds to a dilepton + MET.
A key discriminating variable in this analysis is $m_{T2}$~\cite{Lester:1999tx}, which is typically employed in searches with pair-produced SUSY particles.

This recast is implemented in \texttt{CheckMATE2} at $\sqrt{s}$ = 14 TeV and with $\mathcal{L}$ = 3000 fb$^{-1}$, considering the charginos $\tilde{\chi}^{\pm}_1$, the next-to-lightest neutralino $\tilde{\chi}^{0}_2$, and the sleptons $\tilde{l}$ direct production, with the lightest neutralino $\tilde{\chi}^0_1$ behaving as MET and additionally two leptons in the final state.
The leptons could be of the same or different flavor (SF/DF); however, considering that the RPV signal studied in our work comes from slepton direct pair production and the only non-zero RPV coupling is $\lambda^{\prime}_{111}$, we confine ourselves to the scenario with 2 electrons of opposite signs and large MET.

Candidate events are required to have two opposite-sign electrons with $p_T > 10$ GeV and $|\eta| < 2.47$, and satisfy ``medium'' selection criteria described in Ref.~\cite{ATLAS:2011len}.
The signal electrons should satisfy ``tight'' selection cuts~\cite{ATLAS:2011len} and be isolated with criteria detailed in~\cite{ATLAS:2019lff}. Jets are also reconstructed and are required to have $p_T > 20 $ GeV and $|\eta| < 2.4$. Jets within $|\eta| < 4.9$ are considered in the reconstruction of missing transverse momentum, $p^{\text{miss}}_T$ or MET~\cite{ATLAS:2019lff}, defined as the negative sum of the transverse momenta of all electron candidates with $p_T>10$ GeV and $p_T>20$ GeV jets.

\subsection{Cuts and Signal Regions}

The following cuts are applied to the selected events:
\begin{itemize}
\item $p_T > 35 $ GeV for the first (ranked by $p_T$ in descending order) electron and $p_T > 20$ GeV for the opposite-sign electron.
\item The dilepton invariant mass $m_{ll}$ must be greater than 20 GeV.
\item For our recast, we do not require any jets.\footnote{The full recast within \texttt{CheckMATE2} targets also pair production and decays of charginos, where jet cuts are used.}
\item The absolute difference $|m_{ll}-m_{Z}|$ must be greater than 10 GeV.
\end{itemize}
The SRs are defined by the discriminating variable in the analysis, the ``stransverse mass'' $m_{T2}$, as 
\begin{equation*}
    m^2_{T2} = \min_{\slashed{p_1 }+\slashed{p_2} = \slashed{p_T}}  \big [ \max\{ m^2_T(p_{Tl}, \slashed{p_1}),m^2_T(p_{T\bar l},\slashed{p_2})\}\big]
\end{equation*}
with $m_T$ the transverse mass, $l$ and $\bar l$ are the two opposite-sign leptons, and $\slashed{p_1}$, $\slashed{p_2}$ are the  MET associated with the neutralinos, with $\slashed{p_T}$ being the total MET.

The $m_{T2}$ variable is a generalization of $m_T$, where we have a final state with two invisible objects and two or more visible objects.
In our case, the invisible object corresponds to the neutralinos that escape the inner detector.
We compute $m_{T2}$ from the files in the Oxbridge Kinetics Library~\cite{MT2-StransverseMasss}.
The signal regions are defined as $m^x_{T2}> x$, with $x = 200, 250, 270,$ or $300$ GeV with a fixed value of the neutralino mass of 100 GeV.

\subsection{$m_{T2} $ distributions for our signal}

\begin{figure}[H]
    \centering
    \includegraphics[width=0.495\textwidth]{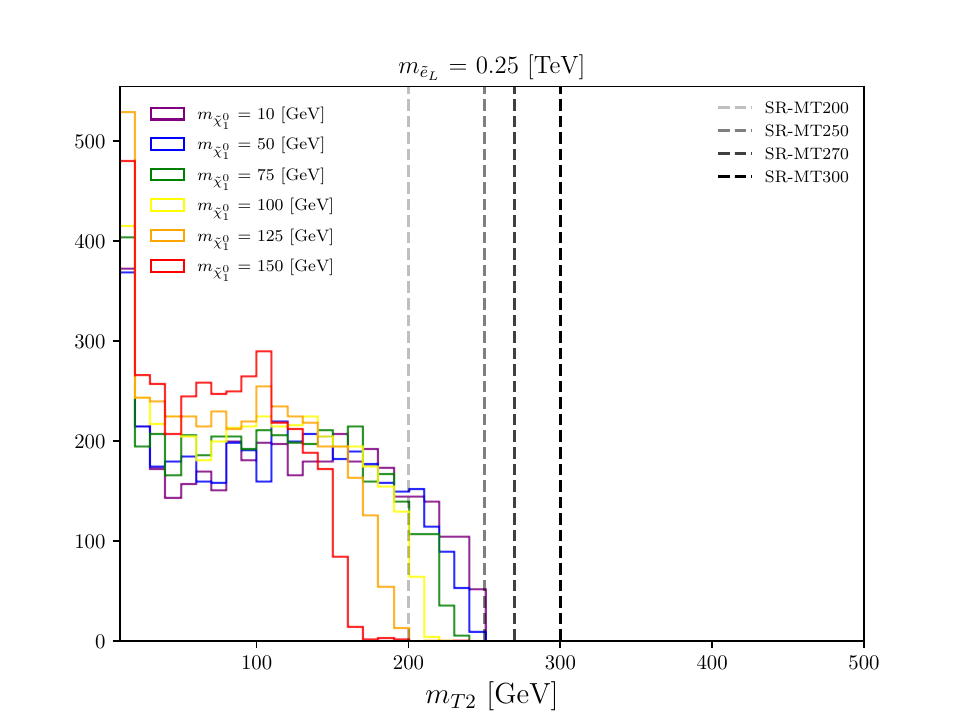}
    \includegraphics[width=0.495\textwidth]{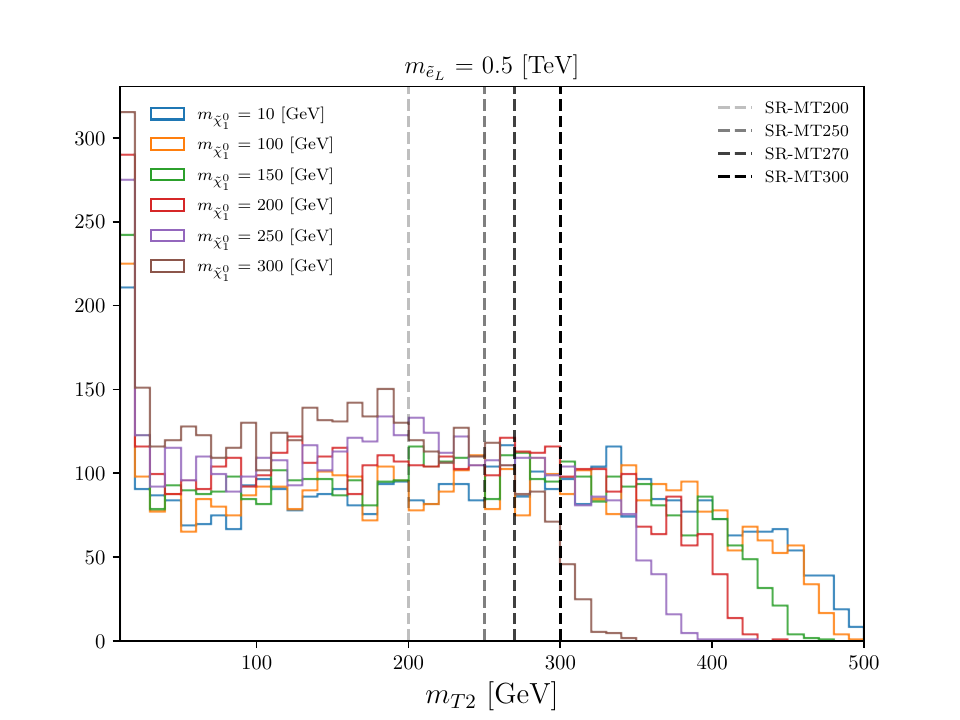}
    \caption{Distributions of $m_{T2} $ for several values of the neutralino mass and fixed choices for the selectron mass $m_{\tilde{e}_L}$ of 0.25 TeV (left) and 0.5 TeV (right).
    The dotted lines represent the SRs defined in the \texttt{CheckMATE2} reinterpretation.}
    \label{fig:mt2_dist}
\end{figure}

Our limits from the ATLAS ``2-leptons + MET'' search can be interpreted on the basis of the behavior of $m_{T2}$ for different values of $m_{\tilde{\chi}^0_1}$ and fixed values of $\lambda^{\prime}_{111}$ = 10$^{-7}$ and $m_{\tilde{e}_L}= 0.25, 0.5$ TeV.
In figure~\ref{fig:mt2_dist}, we show the stransverse mass for selected benchmarks.
The doted gray lines indicate the cut that defines each signal region.

As detailed in Ref.~\cite{Lester:1999tx}, we check that the maximum value of $m_{T2}$ is equal or less than the slepton mass.
We see that for $m_{\tilde{\chi}^0_1} = 10$ GeV, at the end of the distribution, the maximum value of $m_{T2}$ coincides with the actual value of the slepton mass.
For larger values of $m_{\tilde{\chi}^0_1}$, the end of the distribution is shifted towards the left, making all the SR's less sensitive.
This explains the big sharp drop in efficiency observed in the recast curve in Figure~\ref{fig:limits_final} at $m_{\neut}\approx 200$ GeV (for an slepton mass of $0.5$ TeV.).

\begin{figure}[H]
    \centering
    
\end{figure}

\bibliography{main}
\bibliographystyle{JHEP}
\end{document}